\documentclass{ccjnl}

\title{

Interference Cancellation Based Neural Receiver for Superimposed Pilot in Multi-Layer Transmission
}
\author{Han Xiao\inst{1}, Wenqiang Tian\inst{1,*}, Shi Jin\inst{2}, Wendong Liu\inst{1}, Jia Shen\inst{1}, Zhihua Shi\inst{1} and Zhi Zhang\inst{1}
\corinfo{tianwenqiang@oppo.com}
}
\receiveddate{}
\reviseddate{}
\Editor{}

\address[1]{Dept. of Standards Research, OPPO, Beijing, 100026, China}
\address[2]{National Mobile Communications Research Laboratory, Southeast University, Nanjing 211189, China}

\begin{document}
\maketitle

\begin{abstract}
In this paper, an interference cancellation based neural receiver for superimposed pilot (SIP) in multi-layer transmission is proposed, where the data and pilot are non-orthogonally superimposed in the same time-frequency resource. Specifically, to deal with the intra-layer and inter-layer interference of SIP under multi-layer transmission, the interference cancellation with superimposed symbol aided channel estimation is leveraged in the neural receiver, accompanied by the pre-design of pilot code-division orthogonal mechanism at transmitter. In addition, to address the complexity issue for inter-vendor collaboration and the generalization problem in practical deployments, respectively, this paper also provides a fixed SIP (F-SIP) design based on constant pilot power ratio and scalable mechanisms for different modulation and coding schemes (MCSs) and transmission layers. Simulation results demonstrate the superiority of the proposed schemes on the performance of block error rate and throughput compared with existing counterparts.
\end{abstract}

\section{Introduction}
\label{introductionSectionI}
The accurate channel estimation is a key issue of wireless communication systems, which can be achieved through various kinds of pilots in the fifth generation (5G) new radio (NR) system \cite{TS0000,5Ghai,series2015imt}, such as demodulation reference signal (DMRS), channel state information reference signal (CSI-RS) and sounding reference signal (SRS). Towards the sixth generation (6G) \cite{wang2023road,alsabah20216g}, we can expect to see greater advancements in massive multiple input multiple output (MIMO), hybrid beamforming and high-speed scenarios, as well as an increased focus on vertical applications such as sensing and positioning. These will undoubtedly lead to a further growing demand for diverse kinds of pilots, which may result in increased competition for air interface wireless resources between data and pilot transmission. 

In the 5G NR system \cite{TS0000}, pilot design has been standardized as a series of pre-defined patterns and sequences, which fail to consider the implicit channel characteristics of specific scenarios. Recently, deep learning (DL) based methods for air interface enhancement  show great potential in system performance improvement \cite{guo2022overview, hoydis2021toward,TR01111}. Specifically, DL based pilot design including sequence \cite{ma2020data,sohrabi2021deep,chun2019deep,xu2019deep} and pattern \cite{soltani2020pilot,mashhadi2021pruning} with corresponding neural network (NN) receiver are proposed to learn the optimal pilot and the receiver corresponding to specific channel characteristics. However, pilot in above solutions is allocated orthogonally to the data which results in considerable pilot overhead so that a loss of spectral efficiency. A non-orthogonal solution namely superimposed pilot (SIP) \cite{hoeher1999channel} allocates the pilot and data in the same time and frequency resource grids to alleviate the pilot overhead problem, where corresponding pilot power distribution and neural receiver can be jointly trained in an end-to-end manner \cite{aoudia2021end}.


Despite the great throughput performance with reduced pilot overhead, Exisiting DL based SIP \cite{aoudia2021end} also suffers some drawbacks from the perspective of multi-layer transmission and practical deployment. Firstly, multi-layer transmission by precoding uses multiple transmit and receive antennas to simultaneously send multiple data streams, significantly enhancing throughput and making it crucial for advanced standards like 5G and beyond. However, among exisiting DL based SIP methods, there is no consideration dealing with the more serious intra-layer and inter-layer interference caused by SIP in multi-layer transmission, which results in performance loss and calls for brand-new architecture at both transmitter and receiver. Secondly, trainable parameters are at both base station (BS) and user equipment (UE), where the two-sided framework brings much more complexity to inter-vendor training collaboration, e.g. data collection, model training, monitoring, and other model life cycle management issues \cite{chen20235g}. Thirdly, the generalization over different configurations, such as number of transmission layers and modulation and coding scheme (MCS) \cite{TS0002}, is also ignored. 

In this paper, an interference cancellation based neural receiver for SIP in multi-layer transmission is proposed, which involves the innovative design of multiple mechanisms to face the challenges of multi-layer transmission and practical deployment. The main contributions of this artical are summarized as follows.

\begin{itemize}
\item 
To deal with the intra-layer and inter-layer interference of SIP in multi-layer transmission, the interference cancellation with superimposed symbol aided channel estimation is leveraged in the neural receiver, accompanied by the pre-design of pilot code-division orthogonal mechanism at transmitter.
\item 
Considering the practical deployment and standardization, a fixed SIP (F-SIP) based on constant pilot power ratio is designed where the realized one-sided model simplifies the inter-vendor collaboration.
\item 
To address the generalization problem in practical deployment, the scalable mechanisms for different modulation and coding schemes (MCSs) and transmission layers are also proposed, where one same model can work effectively in different MCS and layer configurations.
\item 
Various kinds of simulation results are provided to demonstrate the superiority of the proposed scheme on the performance of block error rate (BLER) and throughput compared with existing counterparts. These abundant simulations are performed with 3rd Generation Partnership Project (3GPP) link level channels, which may hopefully provide some referable insights for 3GPP discussions in the future.
\end{itemize}

The rest of this paper is organized as follows. The system model and existing pilot solutions are introduced in Section \ref{systemdes}. The proposed scheme which involves the innovative design of multiple mechanisms at the transmitter and receiver is proposed in Section \ref{proposed_framework_section}. Numerical experiments are provided in Section \ref{simulation_section}, and conclusions are given in Section \ref{conclusion_section}.

\section{System Description}
\label{systemdes}
\subsection{System Model}
\label{systemmodel}
We consider a typical downlink MIMO system with $N_\textrm{t}$ transmit antennas at BS and $N_\textrm{r}$ receive antennas at UE, where $S$ subcarriers with $T$ consecutive orthogonal frequency division multiplexing (OFDM) symbols are allocated. Specifically, since we mainly focus on multi-layer transmission, the equivalent downlink channel tensor after precoding in frequency domain can be denoted as $\mathbf{H} \in \mathbb{C}^{S \times T  \times L  \times N_\textrm{r}}$, where $L$ denote the number of layers. Received signal for the $r$th receive antenna can be expressed as
\begin{equation}
\mathbf{Y}_{r} = \sum_{l=1}^{L} \mathbf{H}_{r,l}\circ\mathbf{X}_{l} + \mathbf{N}_{r}
\label{receivedsignal}
\end{equation}
where $\mathbf{Y}_{r} \in \mathbb{C}^{S \times T}$ denote the received signal, $1 \leq r \leq N_\textrm{r}$ and $1 \leq l \leq L$ are the receive antenna index and layer index, respectively. $\circ$ denotes the Hadamard product, $\mathbf{H}_{r,l} \in \mathbb{C}^{S \times T}$ is a slice of tensor $\mathbf{H}$ and denotes the equivalent channel for the $r$th receive antenna and $l$th layer. $\mathbf{N}_{r} \in \mathbb{C}^{S \times T}$ is the corresponding additive white complex Gaussian noise with variance of $\sigma^2$ per element according to signal to noise ratio SNR $=10\log_{10}{(\mathbb{E}\{\sum_{l=1}^{L}|x_{l,s,t}|^2\} / \sigma^2)}$. $\mathbf{X}_{l}\in \mathbb{C}^{S \times T}$ is the matrix of transmitted symbols for the $l$th layer, which is capable of carrying data, pilot namely DMRS in 5G NR or DL based orthogonal pilot, or superimposed symbols from pilot and data as introduced later. For all schemes, the transmitted symbols are assumed to have an average energy equal to one, i.e., $\mathbb{E}\{\sum_{l=1}^{L}|x_{l,s,t}|^2\}=1$, where $1\leq s \leq S$ and $1\leq t \leq T$.

\subsection{DMRS in 5G NR}
\label{DMRS5G}
\begin{figure}[tb]
\centering
\includegraphics[scale=0.86]{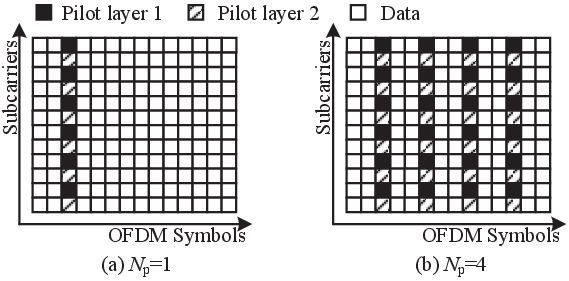}
\caption{Pilot patterns from 5G NR taking the number of layers $L=2$ as an example.}
\label{5Gpattern}
\end{figure}

In this subsection, typical existing pilot solution of demodulation reference signal (DMRS) standardized in 5G NR \cite{TS0000} is introduced, wherein the pilots and data symbols are orthogonally allocated on different resource elements (REs). As shown in Fig. \ref{5Gpattern}, two basic pilot patterns with number of OFDM symbols carrying pilot per slot $N_{\textrm{p}}=1$ and $N_{\textrm{p}}=4$ are designed for lower and higher speed, respectively.  Meanwhile, pilots between different layers are designed orthogonally by frequency-division multiplexing (FDM) and code-division multiplexing (CDM).  

Based on the pilots with pre-defined patterns, the legacy receiver performs channel estimation and data detection with some linear algorithms, such as linear minimum mean square error (LMMSE). Obviously, this kind of orthogonal pilot patterns bring inevitable overhead. In addition, the pattern switching for different scenarios also lead to cumbersome signaling exchange between BS and UE. Moreover, empirically designed pattern fail to consider the implicit characteristics of increasingly complex channel scenarios. These bottlenecks may result in considerable performance loss of throughput. 

\subsection{Pilot based on DL}
\label{DLpilot}
DL-based pilot design shows significant improvements compared with traditional solution in 5G NR. For DL-based orthogonal pilot design, the trainable parameters $\Phi$ and  $\Theta$ are equipped on the transmitter $g(\cdot;\Phi)$ and receiver $f(\cdot;\Theta)$, respectively, where $\Phi$ configures the pilot sequence \cite{ma2020data} or pilot pattern \cite{soltani2020pilot}, and $\Theta$ underpins neural receiver. The parameters $\Phi$ and  $\Theta$ are jointly trained through an end-to-end manner, i.e.,
\begin{equation}
\label{SIPproblem1}
\begin{split}
\min_{\Phi, \Theta}\ \mathcal{L}_{\textrm{bce}}(\mathbf{B},f(h(g(\mathbf{B};\Phi));\Theta)
\end{split}
\end{equation}
where $\mathbf{B} \in \{0,1\}^{S \times T \times M}$ denotes the original encoded information bits, $g(\mathbf{B};\Phi)$ denotes the transmitting signal, $h(\cdot)$ denotes the process of passing channel, and $f(\mathbf{Y};\Theta)$ represents the recovered bits or corresponding log-likelihood ratio (LLR), respectively. $M$ is the number of bits per symbol according to the modulation order, $\mathbf{Y} = h(g(\mathbf{B};\Phi)) \in \mathbb{C}^{S \times T \times N_{\textrm{r}}}$ is the received signal, and $\mathcal{L}_{\textrm{bce}}$ denotes the binary crossentropy loss function. Obviously, orthogonal pilot in above solutions bring cumbersome signaling for pattern switching and inevitable overhead for pilot allocation. As for non-orthogonal DL-based solution \cite{aoudia2021end} where pilot and data are superimposed and the parameters $\Phi$ configures pilot power ratio, there are still non-negligible challenges in multi-layer transmission. More difficult than non-orthogonal multiple access (NOMA) \cite{mohsan2023survey} problem which only introduces the inter-user data interference, the SIP suffers not only from inter-layer data interference but also from intra-layer pilot and data interference. Moreover, the issues of the complexity of two-sided model and generalization of MCS and number of layers mentioned in Section \ref{introductionSectionI} also need to be addressed.

\section{Proposed Schemes}
\label{proposed_framework_section}

In this section, the motivation of designing the proposed schemes is first discussed. Then the proposed mechanisms at transmitter and receiver are introduced. Finally the total framework is formulated by combining all proposed mechanisms.

\subsection{Motivation}
\label{motivationsection}
\subsubsection{Challenge of SIP in multi-layer transmission}
In single-layer SIP transmission \cite{aoudia2021end}, the neural receiver can handle the interference of non-orthogonal pilot and data well. However, as the number of transmission layers increases, it introduces new challenge of intra-layer and inter-layer interference that is difficult for existing neural receivers to cope with. In more detial, accurate channel estimation is required to mitigate the inter-layer interference exploiting the low correlation of channels of different layers brought by the precoding process. Instead, less intra-layer and inter-layer interference to pilot is also required for accurate channel estimation. Increasing the power of the pilot can reduce intra-layer interference, yet it is followed by a reduction in the equivalent SNR of data. These create intractable contradictions and call for novel design of the framework for solving intra-layer and inter-layer interference challenge under multi-layer transmission of SIP. 
\subsubsection{Challenge of SIP in practical deployment}
Considering the NN model in practical wireless communication systems, it is essential to employ appropriate parameter training to adapt the model to different transmission conditions and deploy it with low latency.  However, to implement SIP, trainable paramenters of power ratio and NN model are at the both transmitter and receiver, respectively. This two-sided model brings cumbersome inter-vender collaboration such as data collection, model training, updating and switching \cite{chen20235g}. Moreover, appropriate structure design for different system configuration also brings the problem of generalization. Specifically, distinct configuration such as layers and MCSs can lead to varying dimensions of the neural receiver inputs and outputs so that the different model structure. Thus, it cannot simply utilize the mixed dataset for model training to achieve structure generalization. Consequently, it is imperative to devise an efficient one-sided neural receiver capable of addressing the model structure generalization challenge, rather than relying on extensive model life cycle management to ensure model deployment and application.

\subsection{Pre-design at Transmitter}
Before introducing the interference cancellation based neural receiver, the pre-design at transmitter including intra-layer non-orthogonal F-SIP and inter-layer orthogonal code-division pilot are firstly proposed in this section to deal with the challenges of SIP in practical deployment and multi-layer transmission, respectively.
\subsubsection{Intra-Layer Non-Orthogonal Fixed Superimposed Pilot}
Intra-layer non-orthogonal F-SIP is first introduced in this subsection. Different from the orthogonal pilot patterns in 5G NR, the pilot and data symbols in proposed F-SIP are non-orthogonally superimposed in power domain. Different from the exising two-sided model of SIP solution with trainable paramenter at both transmisster and receiver, a fixed power allocation ratio $0<\alpha<1$ is pre-set at transmitter. Then the transmitted matrix $\mathbf{X}_{l}$ in (\ref{receivedsignal}) with superimposed symbols for the $l$th layer can be denoted as
\begin{equation}
\mathbf{X}_{l} = \sqrt{1-\alpha}\mathbf{D}_{l} + \sqrt{\alpha}\mathbf{P}_{l}
\label{superimposedsymbol}
\end{equation}
where $\mathbf{D}_{l}\in \mathbb{Q}^{S \times T}$ and $\mathbf{P}_{l}\in \mathbb{C}^{S \times T}$ denotes the data and pilot matrix for the $l$th layer, respectively, $\mathbb{Q}$ denotes the constellation set according to the MCS configuration. Obviously, all REs are assigned with a unified and fixed power ratio $\alpha$, instead of a trainable parameter matrix $\Phi$ with extra model training complexity for two-sided structure. A model management friendly one-sided framework serves as a premise here, and the time-frequency resource overhead of orthogonal pilot can be completely omitted.

\begin{figure}[tb]
\centering
\includegraphics[scale=0.48]{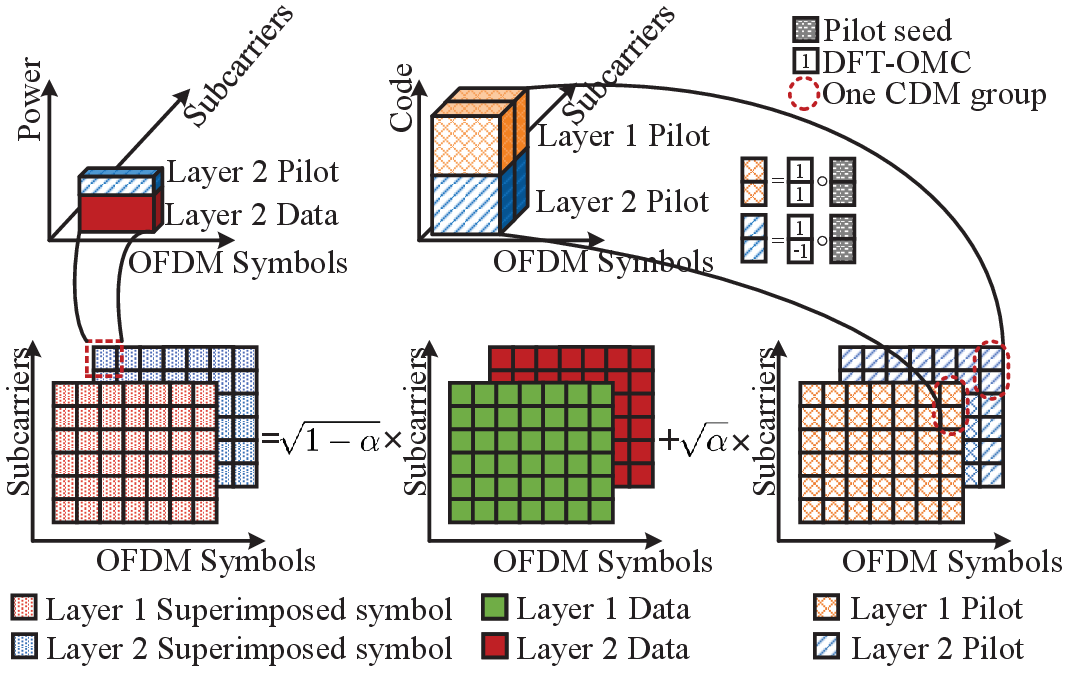}
\caption{Illustration of proposed F-SIP, taking the number of layers $L=2$ as an example for simplicity. Note that the same principle can be extended to the case of $L>2$.}
\label{proposed_1}

\end{figure}

\subsubsection{Inter-Layer Orthogonal Code-Division Pilot}
In this subsection, the F-SIP is further extended to multi-layer transmission, wherein the inter-layer pilot inference introduced by F-SIP should be eliminated. An intuitive way to deal with inter-layer interference is using orthogonal pilots between different layers, such as FDM, time-division multiplexing (TDM) and CDM. However, considering pilots are allocated on orthogonal time and frequency REs in different layers for TDM and FDM, respectively, these two candidates are not suitable for F-SIP transmission where pilots and data are superimposed in all REs. Because the channel estimation based on interpolation in time or frequency domain may result in severer performance loss, especially for high speed or heavy frequency selective scenario. Therefore, CDM is selected in this paper since it can achieve inter-layer pilot orthogonality and meanwhile ensuring all REs can be equipped with pilots.

Specifically, the CDM for F-SIP based multi-layer can be expressed as
\begin{equation}
\|\mathbf{P}_{l}\circ\mathbf{P}_{k}\|_{\rm{F}} = 0, l \neq j
\label{multilayer}
\end{equation}
where the $1 \leq l \leq L$, $1 \leq k \leq L$ are the layer indices and $\|\cdot\|$ denotes the Frobenius norm.  

To satisfy (\ref{multilayer}), all $S \times T$ REs in one layer are devided into $G = S \times T / L$ CDM groups as shown in Fig. \ref{proposed_1}. The pilots of different layers in the same group are distinguished by proposed discrete Fourier transform orthogonal mask code (DFT-OMC), i.e., the pilot sequence can be generated by
\begin{equation}
\mathbf{p}_{l,g} = \hat{p}_{g} \mathbf{c}_{l}
\label{OMC}
\end{equation}
where $\mathbf{p}_{l,g} \in \mathbb{C}^{L \times 1}$ is the vectorized pilot symbols of layer $1 \leq l \leq L$ and group $1 \leq g \leq G$, $\hat{p}_{g}\in \mathbb{P}$ is the pilot seed for group $g$, $\mathbb{P}$ denotes the set of seeded constellation symbols with average power of $1/L$ and zero mean, e.g., binary phase shift keying (BPSK) and quadrature phase shift keying (QPSK). $\mathbf{c}_{l}\in \mathbb{C}^{L \times 1}$ is the DFT-OMC of layer $l$. Furthermore, DFT vectors is utilized to generate $\mathbf{c}_{l}$, , i.e.,
\begin{equation}
\mathbf{c}_{l} = [1, ..., e^{\frac{-j2\pi n(l-1)}{L}},...,e^{\frac{-j2\pi (L-1)(l-1)}{L}}]^{\rm{T}}
\label{OMCvector}
\end{equation}
where $0\leq n \leq L-1$. Using the proposed DFT-OMC, the pilot orthogonality between different layers can be guaranteed with the required power normalization constraint.

\subsection{Interference Cancellation based Neural Receiver}
In this section, a novel neural receiver for F-SIP with multiple enhancing mechanisms is proposed, where the interference cancellation and superimposed symbol aided channel estimation are introduced to cope with the challenge of SIP in multi-layer transmission. The layer and MCS scalable mechanisms are also provided to solve the challenge of SIP in practical deployment.

\begin{figure*}[!t]
\centering
\includegraphics[scale=0.5328]{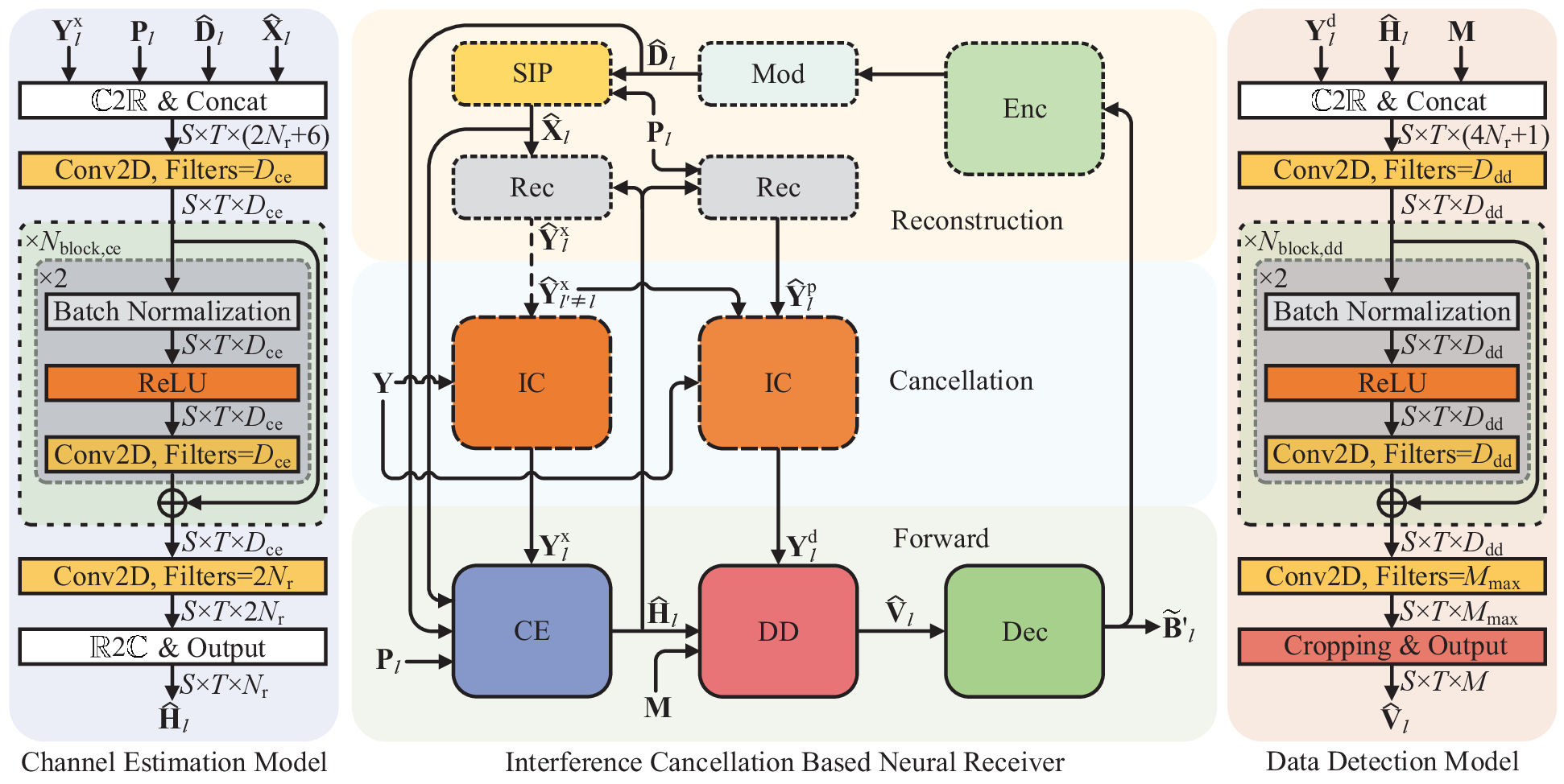}
\caption{Illustration of proposed neural receiver, where the iteration index $i$ is omitted for simplicity. Note that the principle of proposed mechanisms are insensitive to the structure of the feature extraction model for channel estimation and data detection, thus theirs backbones can be implemented flexibly.}
\label{proposed_2}
\end{figure*}

\subsubsection{Interference Cancellation with Superimposed Symbol Aided Channel Estimation}

Next, in order to further handle the inter-layer and intra-layer interference when receiving F-SIP, a receiver with interference cancellation and superimposed symbol aided channel estimation is proposed, inwhich the algorithm includes $V$ outer iterations to realize interference cancellation. Note that the reception for $L$ layers is formulated as $L$ inner iterations in this paper for ease of explanation, which can be parallelized and accelerated by graphics processing unit according to the proposed layer-scalable mechanism as introduced later.  

Fig. \ref{proposed_2} shows the signal processing flow of the proposed receiver, where the IC, DD, CE, Rec modules denote the interference cancellation, data detection, channel estimation and signal reconstruction procedure, and Enc, Dec and Mod modules denote the channel encoding, channel decoding and modulation procedure, respectively. As shown in Fig. \ref{proposed_2}, for $i$th outer iteration and $l$th inner iteration, the inputs of channel estimation model includes the received signal $\mathbf{Y}^{\textrm{x}}_{l,i-1} \in \mathbb{C}^{S \times T \times N_{\textrm{r}}}$ which canceled the reconstructed interference of other $L-1$ layers $\widehat{\mathbf{Y}}^{\textrm{x}}_{l',i-1} \in \mathbb{C}^{S \times T \times N_{\textrm{r}}}, l'\in\{u| 1\leq u \leq L, u \neq l\}$, the reconstructed data of $l$th layer $\widehat{\mathbf{D}}_{l,i-1} \in \mathbb{Q}^{S \times T}$, the reconstructed superimposed symbol of $l$th layer $\widehat{\mathbf{X}}_{l,i-1} \in \mathbb{C}^{S \times T}$ and the pilot of $l$th layer $\mathbf{P}_{l}\in \mathbb{C}^{S \times T}$, inwhich the reconstructed tensors are obtained in $i-1$th iteration and the interference cancellation can be fomulated as 
\begin{equation}
\label{IC_1}
\mathbf{Y}^{\textrm{x}}_{l,i-1} = \mathbf{Y} - \sum_{l'}\widehat{\mathbf{Y}}^{\textrm{x}}_{l',i-1}
\end{equation}
where $\mathbf{Y}  \in \mathbb{C}^{S \times T \times N_{\textrm{r}}}$ denotes the raw received signal concatenating $N_{\textrm{r}}$ received signal $\mathbf{Y}_{r}$ in (\ref{receivedsignal}). Beyond the basic information of pilot components in received signal $\mathbf{Y}^{\textrm{x}}_{l,i-1}$ and pilot $\mathbf{P}_{l}$ for channel estimation, it should be noted that the reconstructed data $\widehat{\mathbf{D}}_{l,i-1}$ and superimposed symbol $\widehat{\mathbf{X}}_{l,i-1}$ are regarded as aided information of `alternative pilots', to make the data components in $\mathbf{Y}^{\textrm{x}}_{l,i-1}$ no longer interfere with channel estimation but can be exploited to enhance the performance of channel estimation. In more detail, with the enhancement of the aided information, the channel can be estimated accroding to three pairs of variables as follows.
\begin{itemize}
\item 
Pilot components in received signal $\mathbf{Y}^{\textrm{x}}_{l,i-1}$ and pilot $\mathbf{P}_{l}$, which is the basic information.
\item 
Data components in received signal $\mathbf{Y}^{\textrm{x}}_{l,i-1}$ and reconstructed data $\widehat{\mathbf{D}}_{l,i-1}$, which is the aided information brought by proposed method.
\item 
Received signal $\mathbf{Y}^{\textrm{x}}_{l,i-1}$ and superimposed symbol $\widehat{\mathbf{X}}_{l,i-1}$, which is also the aided information brought by proposed method.
\end{itemize}
By introducing the aided information, it allows for a lower power ratio of pilot $\alpha$, as long as the channel estimation using only pilot $\mathbf{P}_{l}$ in the first iteration has a certain performance to ignite subsequent iterations since the reconstructed tensors are initialized to all zeros. Data equivalent SNR that is virtually unaffected can hence be guaranteed. Furthermore, even using one of $\widehat{\mathbf{X}}_{l,i-1}$ or $\widehat{\mathbf{D}}_{l,i-1}$ to combine $\mathbf{P}_{l}$ as input provides the same amount of information, all three inputs can facilitate model learning during training phase.

Moreover, the estimated channel of $l$th layer $\widehat{\mathbf{H}}_{l,i} \in \mathbb{C}^{S \times T \times N_\textrm{r}}$ and the received signal $\mathbf{Y}^{\textrm{d}}_{l,i-1} \in \mathbb{C}^{S \times T \times N_{\textrm{r}}}$ are fed into the data detection NN model, where $\mathbf{Y}^{\textrm{d}}_{l,i-1}$ canceled the reconstructed interference of other $L-1$ layers $\widehat{\mathbf{Y}}^{\textrm{x}}_{l',i-1} \in \mathbb{C}^{S \times T \times N_{\textrm{r}}}, l'\in\{u| 1\leq u \leq L, u \neq l\}$ and the reconstructed pilot interference of $l$th layer $\widehat{\mathbf{Y}}^{\textrm{p}}_{l,i-1} \in \mathbb{C}^{S \times T \times N_{\textrm{r}}}$ by using
\begin{equation}
\label{IC_2}
\mathbf{Y}^{\textrm{x}}_{l,i-1} = \mathbf{Y} - \sum_{l'}\widehat{\mathbf{Y}}^{\textrm{x}}_{l',i-1} - \widehat{\mathbf{Y}}^{\textrm{p}}_{l,i-1}
\end{equation}
where $\widehat{\mathbf{Y}}^{\textrm{p}}_{l,i-1}$ is constructed from
\begin{equation}
\label{Rec_interference_2}
\widehat{\mathbf{Y}}^{\textrm{p}}_{l,i-1} = \widehat{\mathbf{H}}_{l,i-1}\circ\mathbf{P}'_{l}
\end{equation}
and $\mathbf{P}'_{l} \in \mathbb{C}^{S \times T \times N_\textrm{r}}$ duplicates the pilot tensor $\mathbf{P}_{l}$ for $N_\textrm{r}$ times. Note that the MCS information $m$ is also the input of the data detection model and is used to achieve MCS generalization, which will be explained in detail later. The model output of LLR tensor $\widehat{\mathbf{V}}_{l,i} \in \mathbb{R}^{S \times T \times M}$ can be further obtained, where $M$ is the number of bits per symbol according to the modulation order indicated by the configured MCS index $m$. In addition to supervision during model training, $\widehat{\mathbf{V}}_{l,i}$ is also exploited for reconstructing the data and superimposed symbol tensor by using
\begin{equation}
\label{Rec_data}
\widehat{\mathbf{D}}_{l,i} = \rm{Mod}(\rm{Enc}(\rm{Dec}(\widehat{\mathbf{V}}_{\it{l,i}})))
\end{equation}
and 
\begin{equation}
\label{Rec_SIP}
\widehat{\mathbf{X}}_{l,i} = \sqrt{1-\alpha}\widehat{\mathbf{D}}_{l,i} + \sqrt{\alpha}\mathbf{P}_{l}
\end{equation}
where $\rm{Dec}(\cdot)$, $\rm{Eec}(\cdot)$ and $\rm{Mod}(\cdot)$ represent the channel decoding, channel encoding and modulation procedure implemented according to the MCS configuration, respectively. $\widehat{\mathbf{B}}'_{l,i} = \rm{Dec}(\widehat{\mathbf{V}}_{l,i})\in \mathbb{R}^{S \times T \times M}$ denotes the received information bits of $l$ layer in $i$th iteration. The interference of $l$th layer for cancellation procedure in $i+1$ iteration can finally calculated  by using
\begin{equation}
\label{Rec_interference_1}
\widehat{\mathbf{Y}}^{\textrm{x}}_{l,i} = \widehat{\mathbf{H}}_{l,i-1}\circ\widehat{\mathbf{X}}'_{l,i}
\end{equation}
and $\widehat{\mathbf{X}}'_{l,i} \in \mathbb{C}^{S \times T \times N_\textrm{r}}$ duplicates the reconstructed superimposed symbol tensor tensor $\widehat{\mathbf{X}}_{l,i}$ for $N_\textrm{r}$ times.

\begin{algorithm}[t]
\caption{Interference Cancellation based Neural Receiver}
\label{alg_proposed}
\begin{algorithmic}
\STATE \hspace*{-5mm} \textbf{Initialization}: $\widehat{\mathbf{Y}}^{\textrm{x}}_{l',0} = \widehat{\mathbf{Y}}^{\textrm{p}}_{l',0} = \mathbf{0}^{S \times T \times N_{\textrm{r}}}$, $\widehat{\mathbf{D}}_{l,0} =\widehat{\mathbf{X}}_{l,0}= \mathbf{0}^{S \times T}$, $1\leq l' \leq L$ and $1\leq l \leq L$;
\STATE \hspace*{-4mm} \textbf{Input}: $\mathbf{Y}$, $\mathbf{P}_{l}$, $m$;
\STATE \hspace*{-4mm} \textbf{Output}: $\widehat{\mathbf{V}}_{l,V}$;
\STATE \hspace*{-4mm} \textbf{for} $i=1 \dots V$ \textbf{do}
\STATE \textbf{for} $l=1 \dots L$ \textbf{do}
\STATE \ \ \ Cancel the data and pilot interference of other $L-1$ layers by using (\ref{IC_1});
\STATE \ \ \ Estimate the channel of the $l$th layer by using the model in section \ref{model_implementation};
\STATE \ \ \ Cancel the data and pilot interference of other $L-1$ layers and pilot interference of the $l$th layer by using (\ref{IC_2});
\STATE \ \ \ Detect the data of the $l$th layer by using the model in section \ref{model_implementation};
\STATE \ \ \ Reconstruct the pilot interference of the $l$th layer by using (\ref{Rec_interference_2});
\STATE \ \ \ Reconstruct the data symbol of the $l$th layer by using (\ref{Rec_data});
\STATE \ \ \ Reconstruct the superimposed symbol of the $l$th layer by using (\ref{Rec_SIP});
\STATE \ \ \ Reconstruct the data and pilot interference of the $l$th layer by using (\ref{Rec_interference_1});
\STATE \textbf{end}
\STATE \hspace*{-5mm} \textbf{end}

\end{algorithmic}
\end{algorithm}

\subsubsection{Layer-Scalable Mechanism}
Under multi-layer transmission, the signals from other layers can be regarded as interference for receiving each target layer, so the problems to be solved in each layer are relatively similar. Therefore, the layer-scalable mechanism is implemented in the proposed receiver, where $L$ layers share same channel estimation and data detection NN models as well as same signal processing flow. The layer scalability can be achieved by proposed layer-common structure since the number of layer only affects the batch size of model inference instead of the inner NN size, where it can be parallelized and accelerated by graphics processing unit conveniently. Meanwhile, since multiple layers share NN structure and parameters, lightweight model are also more friendly to terminal deployment than layer-specific model whose complexity increases with the number of layers increases.

\subsubsection{MCS-Scalable Mechanism}
The MCS generalization is further addressed in this subsection. The inner NN of proposed data detection model is designed according to the maximum number of bits per symbol $M_{\textrm{max}}$ supported by the system, supplemented by the configured MCS index $m$ as auxiliary knowledge, resulting a model structure compatibility with multiple MCSs. Note that the MCS index $m$ is tiled to a tensor $\mathbf{M} \in \{m\}^{S \times T \times 1}$ as input of the NN which facilitates the concatenation of the inputs. After performing feature extraction by NN, the model can proceed a intermediate redundant feature map $\mathbf{V}_{l,i} \in \mathbb{R}^{S \times T \times M_{\textrm{max}}}$. By cropping $\mathbf{V}_{l,i}$ in the third dimension according to $M$, final output of LLR tensor $\widehat{\mathbf{V}}_{l,i} \in \mathbb{R}^{S \times T \times M}$ can be obtained, where $M$ is the number of bits per symbol according to the modulation order indicated by the configured MCS $m$. After collecting all LLR tensors of $L$ layers, it can be fed to the following channel decoder.

\subsubsection{Model Implementation}
\label{model_implementation}

Fig. \ref{proposed_2} shows the NN structure implementing the channel estimation and data detection models. The well-known ResNet \cite{he2016deep} block is utilized wherein double sequential batch normalizations, rectified linear unit (ReLU) activations and two-dimensional convolutional layers (Conv2D) with residual connection are implemented in each block. Since the principle of proposed scalable mechanisms are insensitive to the structure of the feature extraction model, other flexible implementations can be effectively employed such as multi-layer perceptrons mixer \cite{tolstikhin2021mlp} and Transformer \cite{devlin2018bert}. The hyperparameter settings of the model are given in Table \ref{tabSystemParameters} in the simulation part of Section \ref{simulation_section}.

\subsection{Framework of Proposed Scheme}
By combining the above mechanisms, the proposed receiver can be summarized in Algorithm \ref{alg_proposed}. For the sake of simplicity, the Algorithm \ref{alg_proposed} mainly presents stem of the proposed receiver, which helps readers understand the macro framework of the proposed receiver. Finally, the total framework can be formulated as
\begin{equation}
\begin{split}
\min_{\hat{\Theta}_{\textrm{ce}},\hat{\Theta}_{\textrm{dd}}}&\ \frac{1}{V}\sum_{i}^{V}\left\{ \tau\mathcal{L}_{\textrm{bce}}(\widetilde{\mathbf{B}},\widehat{\mathbf{V}}_{i}) + (1-\tau)\mathcal{L}_{\textrm{mse}}(\mathbf{H},\widehat{\mathbf{H}}_{i})\right\}\\
\rm{s.t.} & \ \widehat{\mathbf{V}}_{i}, \widehat{\mathbf{H}}_{i} = \hat{f}(\mathbf{Y},\mathbf{P},L,m; \hat{\Theta}_{\textrm{ce}},\hat{\Theta
}_{\textrm{dd}}))\\
& \ 1 \leq i \leq V
\end{split}
\end{equation}
where $\mathcal{L}_{\textrm{bce}}$ and $\mathcal{L}_{\textrm{mse}}$ denote the binary crossentropy and mean square error loss function, respectively, $\tau$ denotes the weights of loss functions. $\widetilde{\mathbf{B}} \in \{0,1\}^{S \times T \times L \times M}$ and $\mathbf{H} \in \mathbb{C}^{S \times T  \times L  \times N_\textrm{r}}$ represent the original encoded bits and ideal channel, respectively. $\widehat{\mathbf{V}}_{i} \in \mathbb{R}^{S \times T \times L \times M}$ and $\widehat{\mathbf{H}}_{i} \in \in \mathbb{C}^{S \times T  \times L  \times N_\textrm{r}}$ represent the LLR and estimated channel collecting $\widehat{\mathbf{V}}_{l,i}$ and $\widehat{\mathbf{H}}_{l,i}$ of $L$ layers, respectively. $\mathbf{P}\in \mathbb{C}^{S \times T\times L}$ denotes pilot tensor collecting $\mathbf{P}_l$ of $L$ layers. $\hat{f}(\cdot)$,  $\hat{\Theta}_{\textrm{ce}}$ and $\hat{\Theta}_{\textrm{dd}}$ denote the proposed receiver and corresponding NN parameters of channel estimation and data detection model, respectively.

Compared with existing methods, the proposed framework is capable of supporting multi-layer transmission of SIP with practicality and scalability. Challenges mentioned in Section \ref{motivationsection} are well addressed.

\begin{table}[!bp]
\caption{Basic simulation parameters}
\label{tabSystemParameters}
\setlength{\tabcolsep}{4.2mm}{
\begin{tabular}{c|c}
\hline \hline
 Parameter  & Value \\
\hline\hline
Carrier frequency &  4GHz\\
 \hline
Subcarrier spacing  &  30KHz\\
 \hline
PRB number  &  8\\
 \hline
Subcarrier number $S$  &  96\\
 \hline
OFDM symbol number$^{1}$  $T$  &  12\\
 \hline
Tx antennas $N_\textrm{t}$ & 32, 4 \\
 \hline
Rx antennas $N_\textrm{r}$ & 4\\
\hline
 Channel model& CDL\\
 \hline
 Channel coding scheme& LDPC\\
 \hline
 Delay spread $D_{\textrm{s}}$   &  100 ns, 300 ns \\
 \hline
 UE speed $C_{\textrm{ue}}$   &  3-900 km/h \\
 \hline
 Optimizer  &  Adam \\
 \hline
 Training steps  &  2.5 $\times 10^{5}$\\
 \hline
 Training samples number  &  4 $\times 10^{6}$ \\
 \hline
 Weights of loss functions $\tau$  &  0.5 \\
 \hline
 Convolution filter number $D_{\textrm{ce}}$ and $D_{\textrm{dd}}$ &  128 \\
 \hline
 Convolution kernel size  &  3$\times$3 \\
 \hline
 ResNet block number $N_{\textrm{block,ce}}$ and $N_{\textrm{block,dd}}$ &  10 \\
 \hline \hline
\end{tabular}}
\scriptsize{$^{1}$ Two control symbols among 14 symbols in a slot are excluded.}
\end{table}

\section{Simulation Results}
\label{simulation_section}
In this section, numerical results of our proposed F-SIP with scalable neural receiver (Proposed) and two baselines are presented. Specifically, the standardized technology in 5G NR system, i.e., orthogonal pilots of 5G NR in Fig. \ref{5Gpattern} with LMMSE channel estimation and data detection is used as a baseline (Baseline I), wherein the covariance matrix for LMMSE channel estimation is calculated over $10^{5}$ channel samples. It can reflect the gain compared with the existing system design, providing strong simulation result guidance for subsequent application implementation and standardization work. In addition, the state-of-art method from academia  depicted in (\ref{SIPproblem1}) with two-sided model and trainable SIP \cite{aoudia2021end} is also compared as another baseline (Baseline II). The proposed solution provides some improvement methods for a series of problems that the Baseline II does not solve. Therefore, the gain reflected by comparing with this representative SIP solution can illustrate the advancedness of our solution well. The clustered delay line (CDL) channel model is considered here, which has been widely utilized for link-level evaluation in 3GPP \cite{3gpp38901}. Some basic simulation parameters are listed in Table \ref{tabSystemParameters}. The power ratio for F-SIP is set as $\alpha=0.05$ and number of iterations $V=3$ if there is no special declaration. Note that the setting of the hyperparameters of the model in the simulation is based on the trade-off between performance and complexity, which is more in line with practical application. MCS is set as $m=7$ if there is no special declaration, where the modulation scheme is $2^M=16$ quadrature amplitude modulation (QAM) and target coderate is 490/1024. Except for open loop precoder cycling using Type I codebook \cite{TS0002} in high-speed scenario, singular value decomposition (SVD) precoding is used in other scenarios.
During the training phase, each training sample is obtained through the channel sampled from CDL model with random SNR of $-20\sim25$ dB where random SNR in training phase in this paper brings the generalization of SNR of one receiver. Therefore, there is no need to train a specific model for each specific SNR, which is convenient for actual deployment.

\subsection{Effectiveness and Outperformance}
\subsubsection{Comparison in Low Speed Scenario}
\begin{figure}[t]
\centering
\includegraphics[scale=0.48]{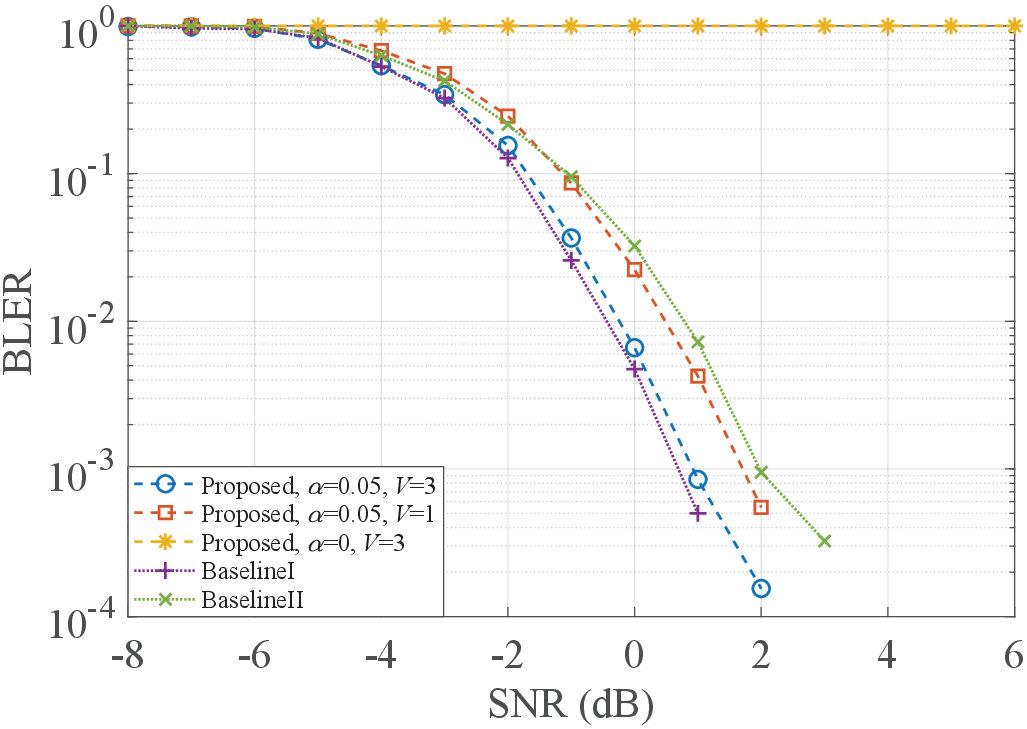}
\caption{BLER performance comparison in low speed scenario ($C_{\textrm{ue}}= 3$ km/h, $D_{\textrm{s}}= 300$ ns, $L=4$ and $N_{\textrm{t}}=32$).}
\label{BLER_1}
\end{figure}

The link-level BLER performance comparison in low speed scenario of CDL-C channel is presented in Fig. \ref{BLER_1}. It can be noticed that the proposed F-SIP with only $\alpha=0.05$ achieve effective BLER performance indicating that the neural receiver can processes the F-SIP well. Specifically, proposed framework outperforms the Baseline II of trainable SIP. This reveals the advantages of proposed aided information and interference cancellation mechanisms under multi-layer transmission while greatly simplifying the system design. The performance at $V=3$ is better than that at $V=1$, indicating the improvement brought by proposed interference cancellation and superimposed symbol aided channel estimation. Moreover, the F-SIP with $\alpha=0$ does not work well, indicating that even with a small power ratio, e.g., $\alpha=0.05$, pilot is necessary for channel estimation in the neural receiver. Proposed F-SIP with $\alpha=0.05$ is comparable with the Baseline I of traditional orthogonal pilot with $N_{\textrm{p}}=1$. This indicats that non-orthogonal pilot bring almost no performance loss, and the proposed method can transmit more effective information bits under same coding rate resulting in better throughput performance, which will be detailed later.

\begin{figure}[t]
\centering
\includegraphics[scale=0.48]{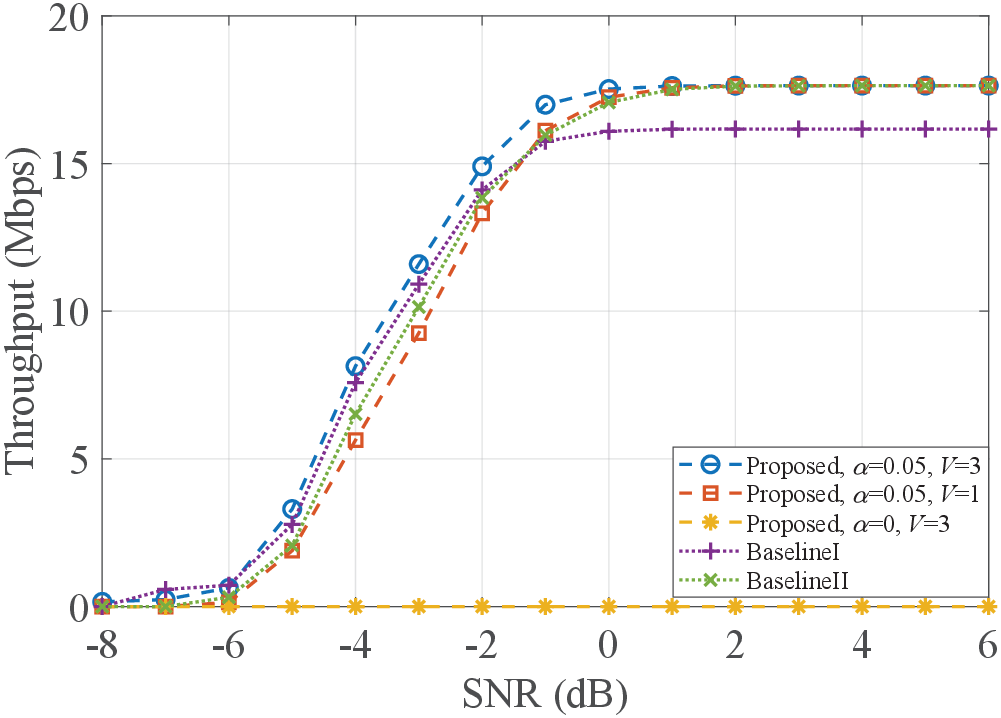}
\caption{Throughtput performance comparison in low speed scenario ($C_{\textrm{ue}}= 3$ km/h, $D_{\textrm{s}}= 300$ ns, $L=4$ and $N_{\textrm{t}}=32$).}
\label{TP_1}
\end{figure}

The throughput comparison in low speed scenario of CDL-C channel is provided in Fig. \ref{TP_1}, where the throughput $R$ is defined as 
\begin{equation}
\begin{split}
R = N_{\textrm{slot}}N_{\textrm{RE}}\Omega\gamma M(1-BLER)
\end{split}
\label{GPcal}
\end{equation}
wherein $N_{\textrm{RE}} = STL$ denotes the number of REs forming a slot, $N_{\textrm{slot}}$ denotes the number of slot per second, $\Omega$ denotes ratio of REs carrying data symbols, $\gamma$ and $M$ are the target coderate and number of bits per symbol according to selected MCS, respectively. For orthogonal pilot patterns in Baseline I, some REs are reserved for pilot transmission. Thus we have $\Omega = 11/12$ for Baseline I in 3km/h, while other methods are with $\Omega = 1$. Obviously, the proposed method with $\alpha = 0.05$ and $V=3$ achieves higher throughput compared with Baseline I. Moreover, we find that our proposed  method with pre-set $\alpha = 0.05$ and $V=3$ can achieve comparable throughput with Baseline II, which demonstrates that one-sided model at only UE side is capable of ensuring the performance with more flexible model management procedure compared with two-sided counterpart.

\begin{figure}[t]
\centering
\includegraphics[scale=0.48]{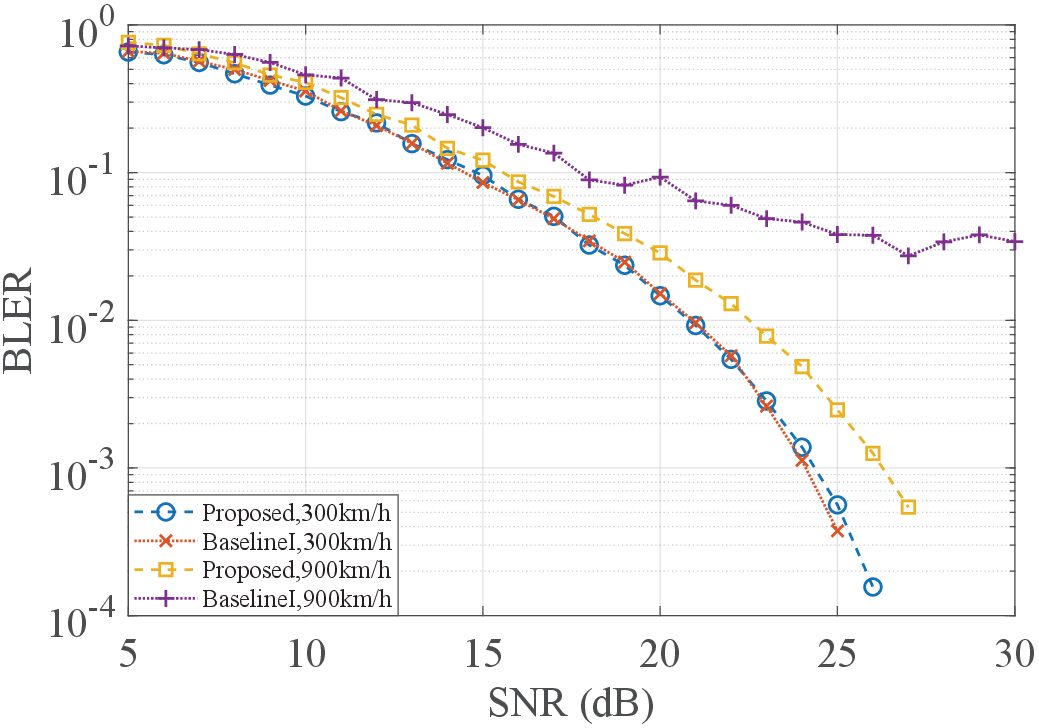}
\caption{BLER performance comparison in high speed scenario ($C_{\textrm{ue}}= 300/900$ km/h, $D_{\textrm{s}}= 100$ ns, $L=2$ and $N_{\textrm{t}}=4$).}
\label{BLER_2}
\end{figure}

\subsubsection{Comparison in High Speed Scenario}
\begin{figure}[t]
\centering
\includegraphics[scale=0.528]{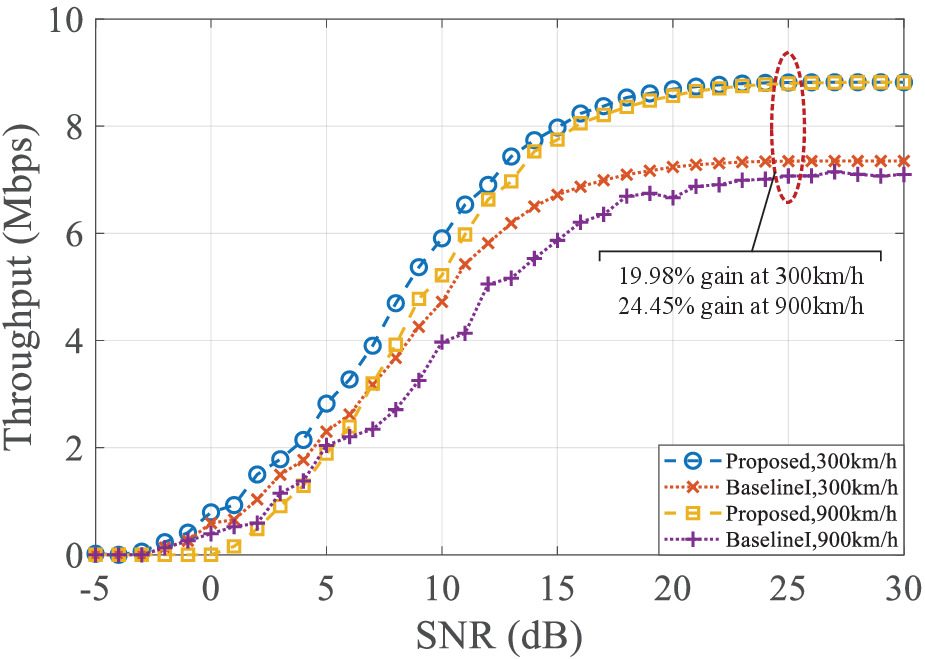}
\caption{Throughtput performance comparison in high speed snenario ($C_{\textrm{ue}}= 300/900$ km/h, $D_{\textrm{s}}= 100$ ns, $L=2$ and $N_{\textrm{t}}=4$).}
\label{TP_2}
\end{figure}

The BLER and throughput performance comparison in high speed scenario with CDL-D extension  channel \cite{FL0009} are depicted in Fig. \ref{BLER_2} and \ref{TP_2}, respectively, where $N_{\textrm{p}} = 4$ for Baseline I is necessarily configured to estimate channels with strong time-varying characteristics. Thus we have $\Omega = 10/12$, while proposed methods are with $\Omega = 1$. Generally, the proposed method achieves higher throughput compared with Baseline I in high speed scenario since the pilot overhead can be avoided. Taking SNR = 25 dB as an example, gains of 19.98\% and 24.45\% can be obtained in scenarios of 300 km/h and 900 km/h, respectively. Moreover, proposed method in extremely high-speed scenarios of 900 km/h has significant performance advantages from the perspective of BLER, as all available time domain and frequency domain resources have pilot distribution. The LMMSE in Baseline I can realize accurate channel estimation based on covariance matrix from $10^5$ channel samples under 300 km/h, while the channel estimation error resulted from interpolation in time domain grows extremely large when 900 km/h. As comparison, our proposed method is capable of performing accurate joint channel estimation and data detection by exploiting the statistical relationship between pilots and data symbols on all REs.

\begin{figure}[t]
\centering
\includegraphics[scale=0.48]{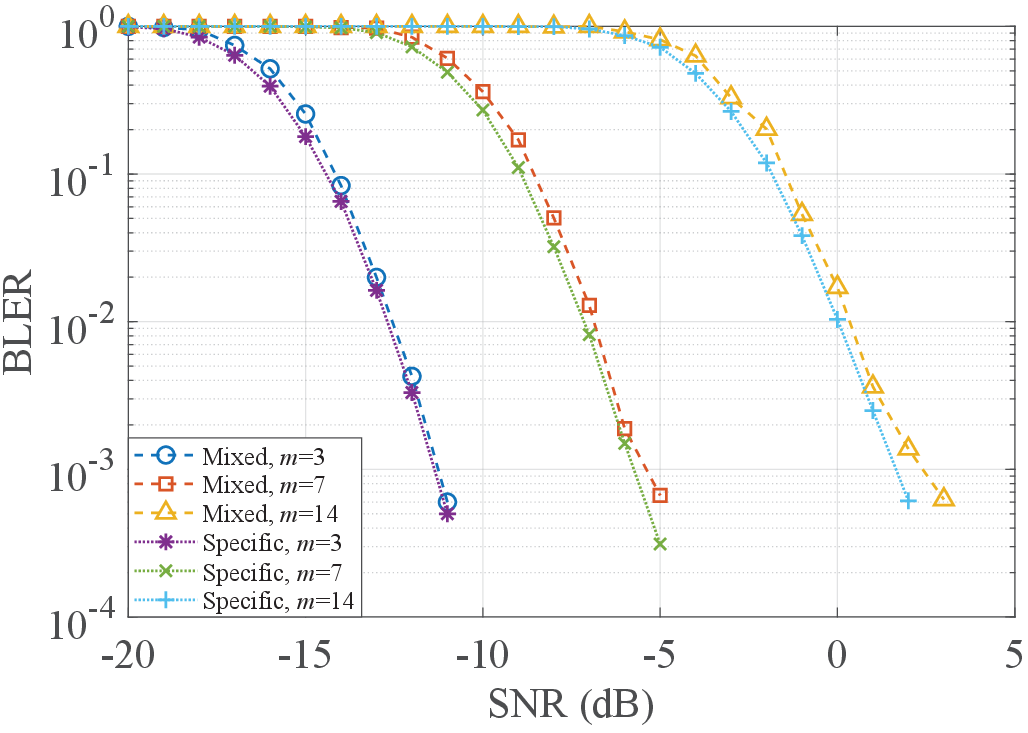}
\caption{Generalization study of proposed scalable neural receiver for different MCSs ($C_{\textrm{ue}}= 3$ km/h, $D_{\textrm{s}}= 300$ ns, $L=2$ and $N_{\textrm{t}}=32$).}
\label{MCSG}
\end{figure}

\subsection{Generalizability and Scalability}
\subsubsection{Scalability for Different MCSs}
The scalability performance on different MCSs of our proposed scalable neural receiver in CDL-C channel is presented in Fig. \ref{MCSG}. Here the MCS $m$=\{3, 7, 14\} are selected with corresponding modulation order as \{QPSK, 16QAM, 64QAM\} and coderate as \{449/1024, 490/1024, 719/1024\}, respectively.  Our proposed scalable neural receiver (Mixed) is trained on the mixed datasets with $m=\{3, 7, 14\}$, with the model implementation of $M_{\textrm{max}}=6$. While the compared specific models (Specific) are implemented and trained on its own single MCS without using proposed scalable mechanisms. It can be noticed that our proposed scalable neural receiver can achieve comparable performance with specific counterparts, which validates its excellent scalability performance on and MCSs.

\begin{figure}[t]
\centering
\includegraphics[scale=0.48]{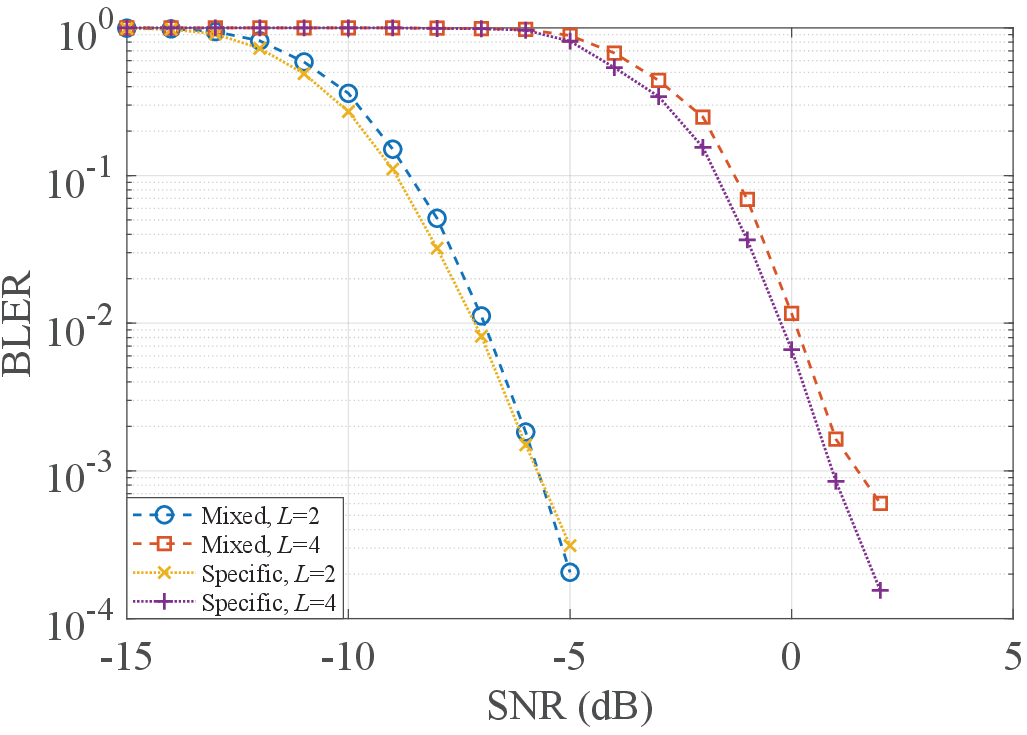}
\caption{Generalization study of proposed scalable neural receiver for different number of layers ($C_{\textrm{ue}}= 3$ km/h, $D_{\textrm{s}}= 300$ ns, $L=2$ and $N_{\textrm{t}}=32$).}
\label{LAYERG}
\end{figure}

\subsubsection{Scalability for Different Number of Layers}
The scalability performance on different number of layers in CDL-C channel is studied in Fig. \ref{LAYERG}, wherethe number of layers is evaluated with $L=\{2, 4\}$. The proposed scalable neural receiver (Mixed) is trained on the mixed datasets with $L=\{2, 4\}$. While the compared specific models (Specific) are implemented and trained on its own single layers. Obviously, comparable performance is obtained, indicating the practical deployment-friendly scalability for different layers can be achieved.

\begin{figure}[t]
\centering
\includegraphics[scale=0.48]{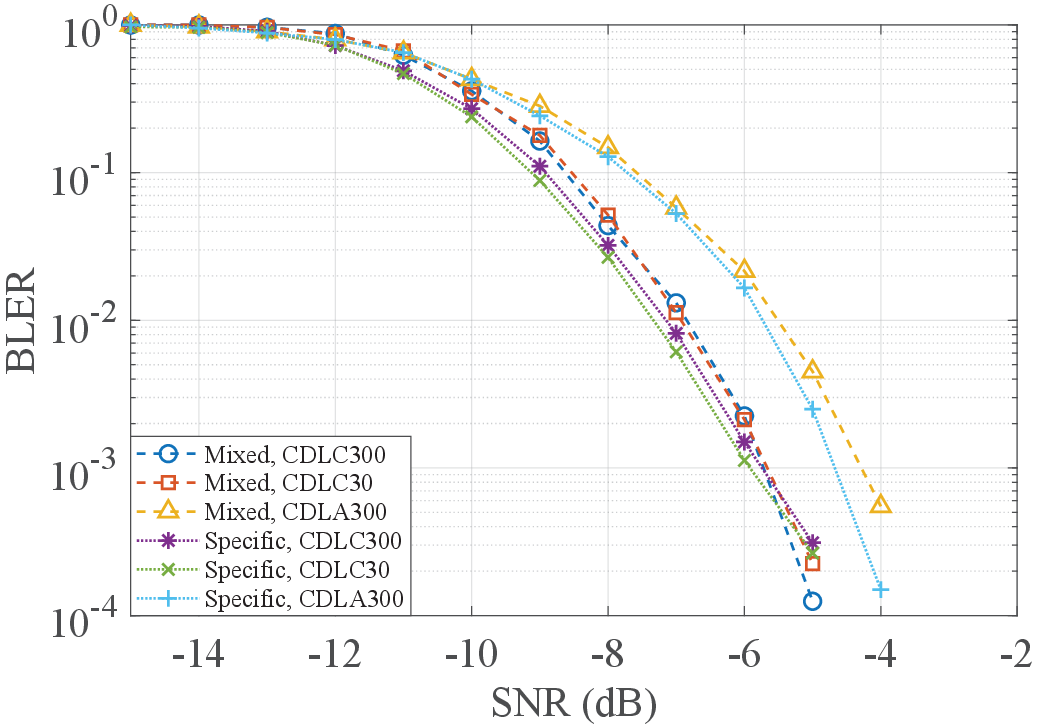}
\caption{Generalization study of proposed scalable neural receiver for different channel models ($C_{\textrm{ue}}= 3$ km/h, $L=2$ and $N_{\textrm{t}}=32$).}
\label{CHANNELG}
\end{figure}

\subsubsection{Generalization for Different Channel Enviroments}
To further explore possibility in practical deployment, generalization study in different channel enviroment is also provided in Fig. \ref{CHANNELG}, where configuring more channel setting to the training dataset can further obtain the generalization of the mixed channel model \cite{liu2021evcsinet}. The proposed neural receiver (Mixed) is trained on the mixed datasets of CDL-A/C with $D_{\textrm{s}} = 30/300$ ns and tested on the corresponding target channel. While the compared specific models (Specific) are trained on the target channel model. It can be seen that proposed receiver still achieves comparable BLER performance, which exhibits the excellent generalization performance in practical deployment when facing different channels.

\subsection{Computational and Storage Complexity}
The computational and storage complexity evaluation is also studied. 
First, from the perspective of simulation, an evaluation of the running time of the model ($M_{\rm{max}}=6$) on single NVIDIA A100 SXM 80 is provided. By processing $1.28 \times 10^{5}$ transport blocks (TB), the averaged computation time for each inner iteration is about 0.5 milliseconds. Moreover, from the perspective of analysis, the complexity of proposed receiver mainly lies in the channel estimation and data detection model which is far beyond the complexity of interference reconstruction and cancellation. Therefore we provide the floating point operations (FLOPs) and trainable parameters evaluation of one iteration. Firstly, the channel estimation model brings 2.9802$\times STL \times $10$^{6}$ FLOPs with 2.9777$\times $10$^{6}$ parameters, where the computational and storage complexity are not affected by $M_{\textrm{max}}$ and $L$, respectively. The complexity of data detection model is provided in Table \ref{tabFLOPs1}. It can also be noticed that the computational complexity increases with the increase of $M_{\textrm{max}}$, $T$, $S$ and $L$, where the influence of $M_{\textrm{max}}$ is relatively negligible compared with the complexity of the model itself. Moreover, not only does $M_{\textrm{max}}$ have a slight impact on storage complexity, but the number of transmission layers $L$ does not have an impact on storage complexity of proposed layer-common structure. While the storage complexity of the the layer-specific model is expanded $L$ times since different layers use different structures and parameters. These imply the feasibility of deployment of proposed scalable receiver.

\begin{table}[!h]
\centering
\caption{Evaluation of FLOPs and number of trainable parameters of data detection model.}
\setlength{\tabcolsep}{4.2mm}{
\begin{tabular}{|c|c|c|c|c|}
\hline
$M_{\textrm{max}}$     & FLOPs ($ \times STL \times 10^{6}$)  & Parameters ($\times 10^{6}$)  \\
 \hline
$2$  & 2.9813 & 2.9788 \\ \hline
$4$  & 2.9836 & 2.9811 \\ \hline
$6$  & 2.9859 & 2.9834 \\ \hline
\end{tabular}}
\label{tabFLOPs1}

\end{table}

\section{Standardization Potential and Prospects}
Starting from 3GPP release 18, the study item of `Artificaial Intelligence / Machine Learning for NR Interface' introduces the DL-based solutions into the physical layer of communication system. Some system design restrictions can be further relaxed using those DL-based approaches, which also makes it possible to explore new forms of reference signal in the subsequent 6G research such as different learnable sequence and pattern as well as introduction of non-orthogonality. 

According to 3GPP's work plan about DL-based solutions from 5G-advanced to 6G, the performance gain, overhead reduction, scenario generalization, storage and computational complexity, life cycle management (LCM) \cite{chen20235g} and potential standardization impact need to be studied. Therefore, the DL-based pilot solutions in existing research also need to address some corresponding challenges to meet the practical requirements and follow a standardized route for 6G, namely i) maintaining the throughput gain in more complex environments such as high-speed scenarios or multi-layer transmission, ii) achieving a lower or zero overhead of reference signal, iii) keeping a lower complexity to adapt to terminal deployment, iv) generalizing to different scenarios or system configurations, and vi) designing the simple framework without cumbersome LCM procedure. 

The solution proposed in this artical involves multiple novel mechanisms design to solve the above challenges from the perspective of throughput, overhead, generalization, scalability, flexibility and complexity, making the SIP compliant with standardization and practical deployment. In future work, it is meaningful to further study the effectiveness and complexity in more practical scenarios before SIP is standardized. These will also bring more diversity and space for the redesign of various reference signal in future 6G intelligent system.

\section{Conclusion}
\label{conclusion_section}
In this paper, an interference cancellation based neural receiver for SIP in multi-layer transmission is proposed, which involves multiple novel mechanisms design to face the challenges of multi-layer transmission and practical deployment. Specifically, considering the intra-layer and inter-layer interference of SIP under multi-layer transmission, the interference cancellation with superimposed symbol aided channel estimation is utilized in the neural receiver, accompanied by the pre-design of pilot code-division orthogonal mechanism at transmitter. Moreover, to deal with the complexity issue for inter-vendor collaboration and the generalization problem for practical deployments, respectively, a fixed SIP (F-SIP) design based on constant pilot power ratio and scalable mechanisms for different modulation and coding schemes (MCSs) and transmission layers are also proposed. Simulation results demonstrate the superiority of the proposed scheme from the perspective of BLER and throughput compared with existing counterparts.

%


\bibliographystyle{gbt7714-numerical}
\bibliography{ref}

%
%

\end{document}